\documentclass[11pt]{article}
\usepackage{blois,epsfig}

\def\Journal#1#2#3#4{{#1} {\bf #2}, #3 (#4)}

\def\PRL{\em Phys. Rev. Lett.}

\def\AA{\em A\&A}

\def\be{\begin{equation}}
\def\ee{\end{equation}}
\def\bea{\begin{eqnarray}}
\def\eea{\end{eqnarray}}

\begin{document}
\vspace*{2cm}
\title{
A Proposal to Search for Transparent Hidden Matter
Using Optical Scintillation
}

\author{ M.~Moniez }

\address{Laboratoire de l'Acc\'{e}l\'{e}rateur Lin\'{e}aire,
{\sc IN2P3-CNRS}, Universit\'e de Paris-Sud, B.P. 34, 91898 Orsay Cedex, France}

\maketitle\abstracts{
It is proposed to search for scintillation of extragalactic sources
through the last unknown baryonic structures.
%due to refraction index of cool molecular 
%Stars twinkle because their light goes through the atmosphere.
%The same phenomenon is expected when the light of extra-galactic stars
%goes through a Galactic -- disk or halo -- refractive medium.
%Because of the large distances involved here, the length and time
%scales of the optical intensity fluctuations resulting from the
%wave distortions are accessible to the current technology.
%In this paper, we discuss the different possible scintillation
%regimes and we focus on the so-called strong diffractive regime
%that is likely to produce large intensity contrasts.
Appropriate observation of the scintillation process
described here should allow one to detect column density
stochastic variations in cool Galactic molecular clouds of order of
$\sim 3\times 10^{-5}\,\mathrm{g/cm^2}$ -- that is
$10^{19}\,\mathrm{molecules/cm^2}$ --
per $\sim 10\,000\,\mathrm{km}$ transverse distance.
}
\section{Introduction}
%Considering the results of baryonic compact massive objects searches
%(\cite{notenoughmachos}; \cite{SMC5ans}; \cite{MACHO}),
%it seems that the only constituent
%that could contribute quite significantly
%to the Galactic baryonic hidden matter is the
%cool molecular hydrogen ($\mathrm{H_2}$).
It has been suggested that a hierarchical structure of cold $\mathrm{H_2}$
could fill the Galactic thick disk \cite{fractal} or halo \cite{Jetzer1}, 
providing a solution for the Galactic dark matter problem.
This gas could form undetectable
``clumpuscules'' of $10\,\mathrm{AU}$ size at the smallest scale,
with a column density of $10^{24-25}\,\mathrm{cm^{-2}}$, and a
surface filling factor less than 1\%.
Light propagation is delayed through such a structure and
the {\it average} transverse gradient of optical path variations is of
order of $1\lambda\ per\ 10\,000\,\mathrm{km}$ at $\lambda=500\,\mathrm{nm}$.
These $\mathrm{H_2}$ clouds could then be detected through their
diffraction and refraction effects on the propagation of the light of
background stars (for a more detailed paper on this proposal
see ( \cite{Moniez})).
\section{Detection mode of extended $\mathrm{H_2}$ clouds: principle}
\begin{figure}
\begin{center}
%\begin{turn}{0}
%\mbox{\epsfig{file=../Scintillation/diffusion-faible-lores.eps,width=7.5cm}}
%\mbox{\epsfig{file=../Scintillation/diffusion-forte-lores.eps,width=7.5cm}}
\mbox{\epsfig{file=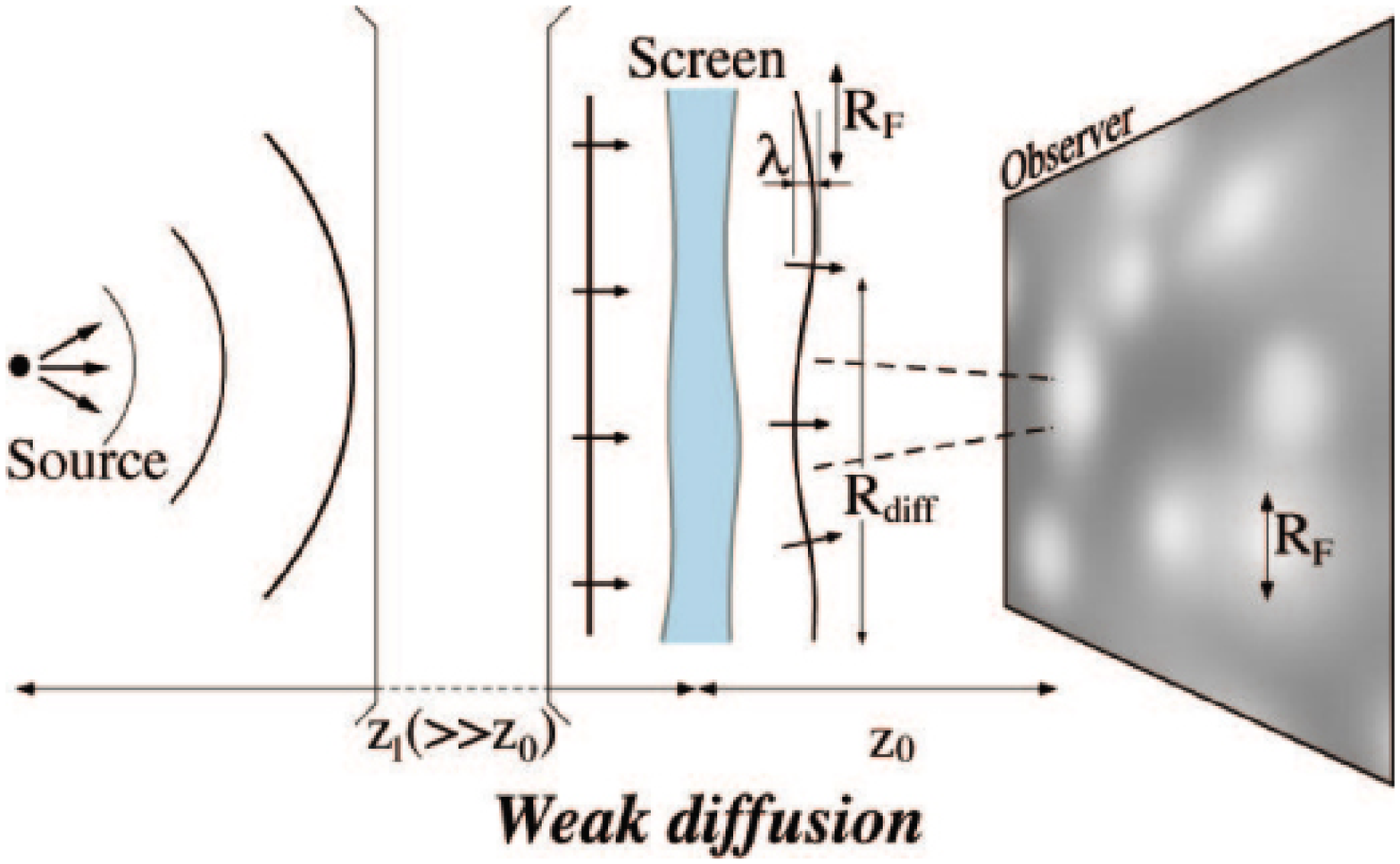,width=7.5cm}}
\mbox{\epsfig{file=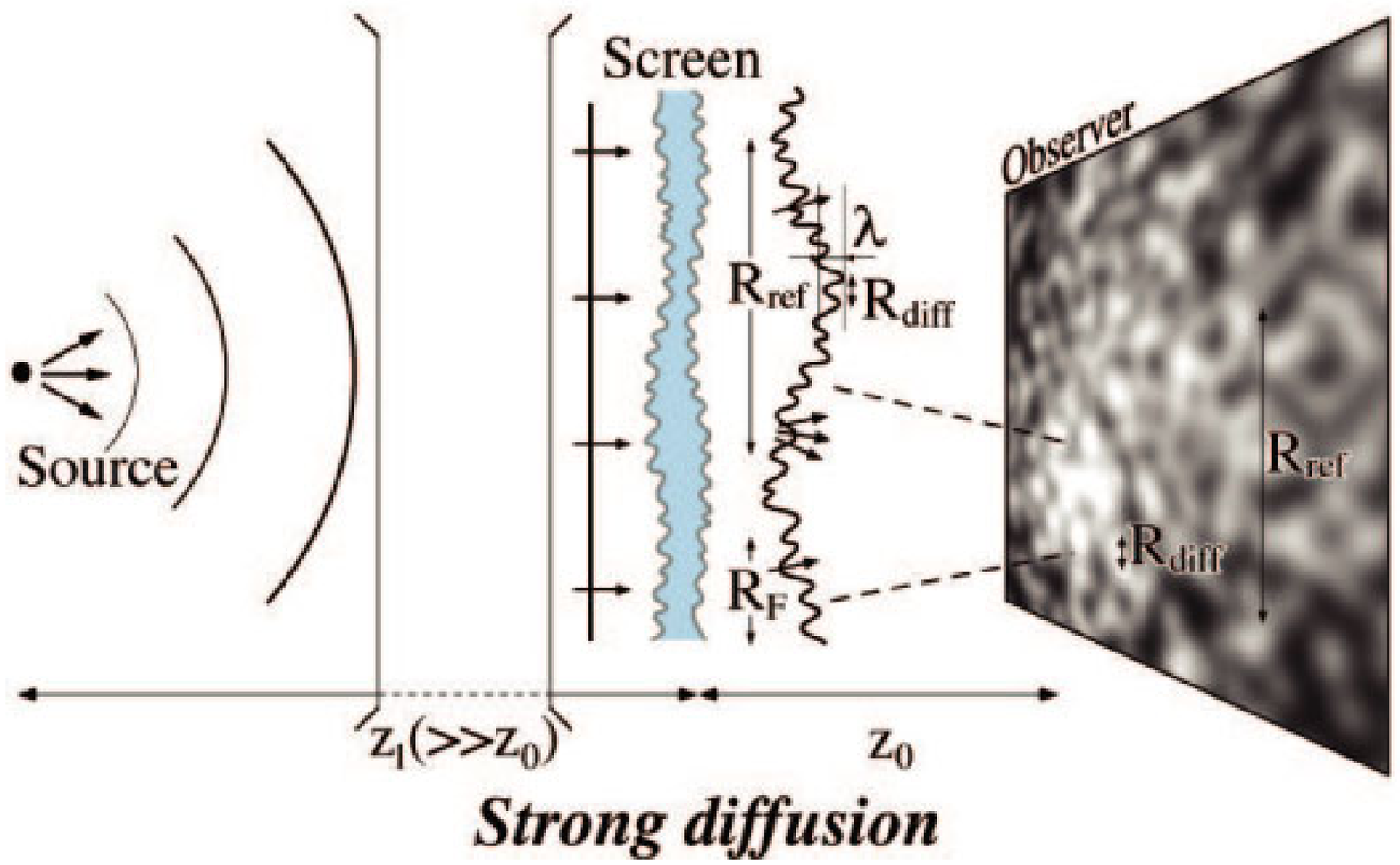,width=7.5cm}}
%\mbox{\epsfig{file=../Scintillation/weak-diffusion.eps,width=7.5cm}}
%\mbox{\epsfig{file=../Scintillation/strong-diffusion.eps,width=7.5cm}}
%\end{turn}
\caption[]{The two scintillation regimes:
Left: $R_{diff}\gg R_F$. The weakly distorted
wavefront produces a weak scintillation at scale $R_F$
in the observer's plane.
Right: $R_{diff}\ll R_F$. The strongly distorted wavefront
produces strong scintillation at scales $R_{diff}$ (diffractive mode)
and $R_{ref}$ (refractive mode).
\label{front}}
\end{center}
\end{figure}

Due to index refraction effects,
an inhomogeneous transparent medium (hereafter called screen)
%at distance $z_0$ from an observer
distorts the wave-fronts of incident electromagnetic waves
(see Fig. \ref{front}).
%In the observer's plane,
%the luminous amplitude of a source located behind a gaseous structure
%(hereafter called screen), results from
%the propagation of the distorted wave-front.
The extra optical path induced by a screen at distance $z_0$
can be described
by a phase delay $\delta(x_1,y_1)$ in the plane transverse
to the observer-source line.
The amplitude $A_0$ in the observer's plane
results from the subsequent propagation, that
is described by the Huygens-Fresnel diffraction theory
\begin{equation}
\label{amplit}
A_0(x_0,y_0)=\frac{Ae^{2i\pi z_0/\lambda}}{2i\pi R_F^2}
\int\!\!\int_{-\infty}^{+\infty}
e^{\frac{2i\pi\delta(x_1,y_1)}{\lambda}}
e^{i\frac{(x_0-x_1)^2+(y_0-y_1)^2}{2 R_F^2}}dx_1dy_1 ,
\end{equation}
where $A$ is the incident amplitude (before the screen),
taken as a constant for a very distant on-axis source,
and $R_F=\sqrt{\lambda z_0/2\pi}$ is the Fresnel radius.
$R_F$ is of order of $1500\,\mathrm{km}$ to $15\,000\,\mathrm{km}$
at $\lambda=500\,\mathrm{nm}$,
for a screen distance $z_0$ between $1\,\mathrm{kpc}$ to $100\,\mathrm{kpc}$.
%This length scale characterizes the $(x_1,y_1)$ domain that contributes
%to the integral (a few Fresnel radii).
The resulting intensity in the observer's plane
is affected by strong interferences
(the so-called speckle) if 
$\delta(x_1,y_1)$ varies stochastically of order of $\lambda$
within the Fresnel radius domain.
This is precisely the same order of magnitude as
the average gradient that characterizes the hypothetic
$\mathrm{H_2}$ structures.

As for radio-astronomy \cite{lyne}$^,$\cite{narayan92},
the stochastic variations of $\delta(x_1,y_1)$ can be characterized
by the diffractive length scale $R_{diff}$, defined as the
transverse separation
for which the root mean square of the optical path difference
is $\lambda/2\pi$.
\begin{itemize}
\item
If $R_{diff}\gg R_F$, the screen is weakly diffusive.
The wavefront is weakly corrugated, producing
weakly contrasted patterns with length scale $R_F$ in the observer's plane
(Fig. \ref{front}).
\item
If $R_{diff}\ll R_F$, the screen is strongly diffusive.
Two modes occur, the diffractive scintillation -- producing strongly
contrasted patterns characterized by the length scale $R_{diff}$ --
and the refractive scintillation -- giving less contrasted patterns and
characterized by the large scale structures of the screen $R_{ref}$ --.
\end{itemize}
We focus here on the strong diffractive mode, which
should produce the most contrasted patterns and is easily predictable.
\section{Basic configurations}
\label{simplecase}
\begin{figure}
\begin{center}
%\mbox{\epsfig{file=../Scintillation/frange-point-moy.eps,width=5.2cm}}
%\mbox{\epsfig{file=../Scintillation/point-coin-moy.eps,width=5.2cm}}
%\mbox{\epsfig{file=../Scintillation/franges-RS1-2-4.eps,width=5.2cm}}
\mbox{\epsfig{file=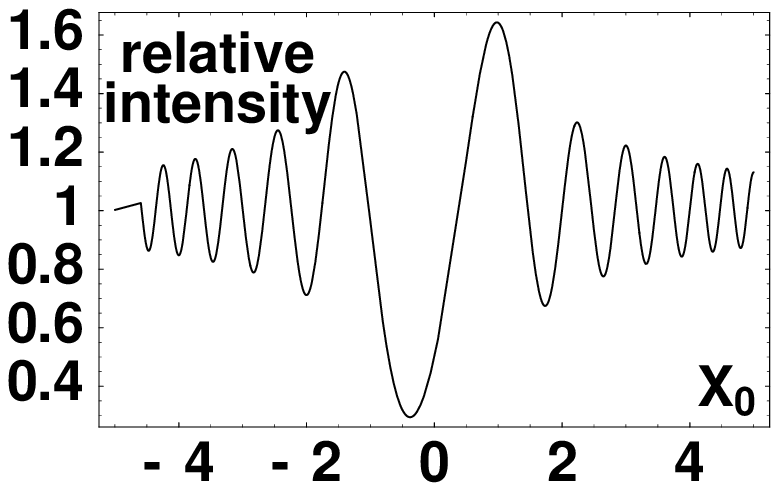,width=5.2cm}}
\mbox{\epsfig{file=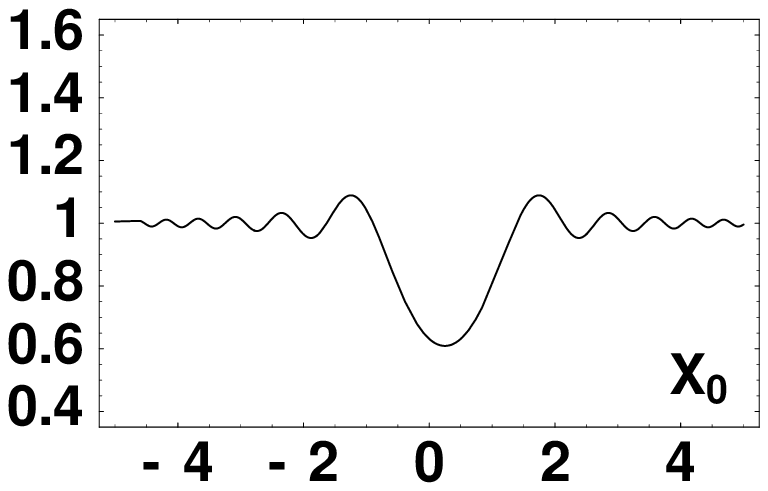,width=5.2cm}}
\mbox{\epsfig{file=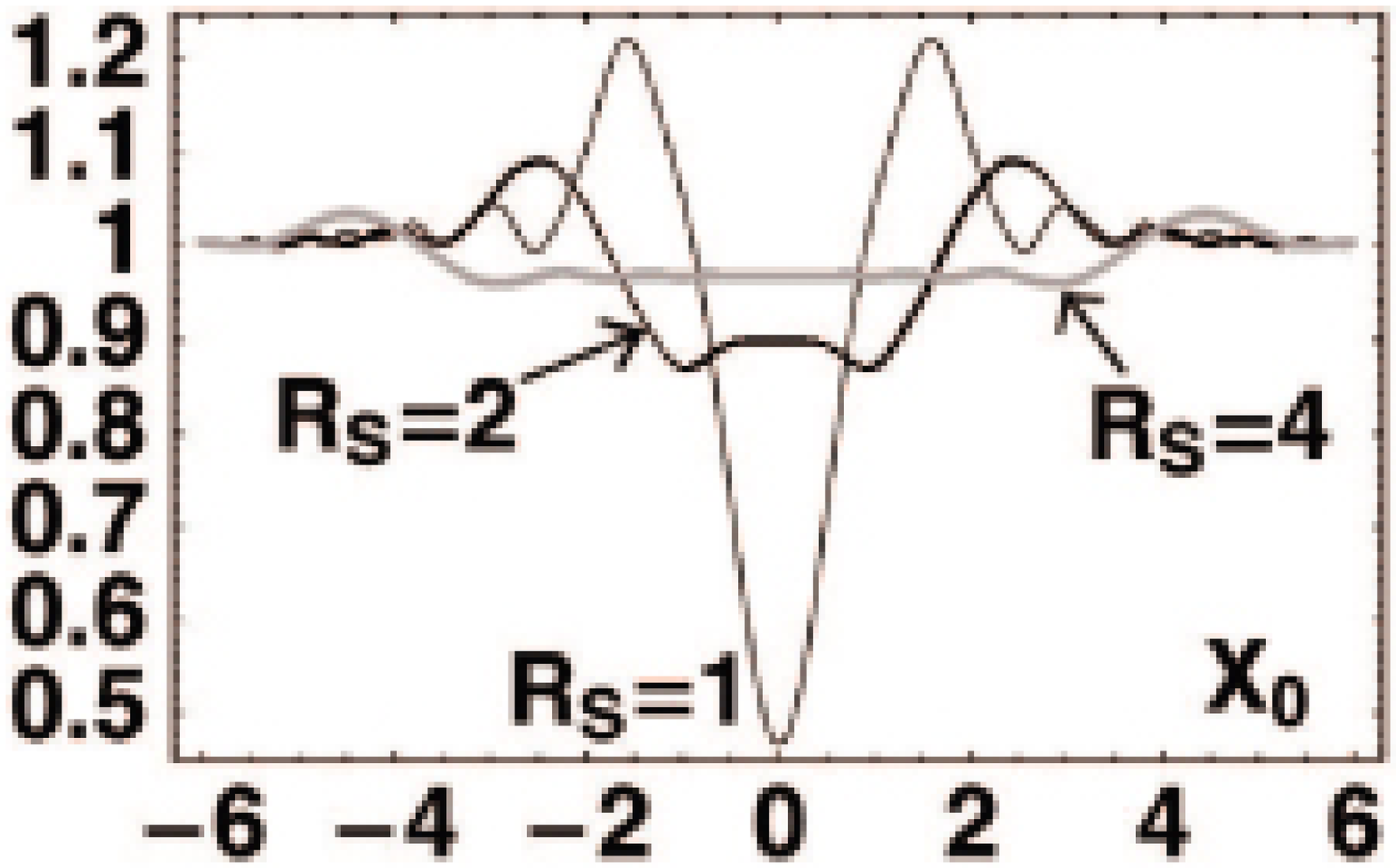,width=5.2cm}}
%\end{turn}
\caption[]{Relative intensity diffraction patterns produced in the
observer's plane, perpendicularly to a width step or prism.
From left to right:
- Step of optical path $\delta=\lambda/4$
- Prism of optical path with $\delta=X_0\times\lambda/2$ for $X_0>0$.
- Step of optical path $\delta=\lambda/2$, for extended disk-sources
of reduced radius $R_S=1$, 2 and 4.
The $X_0-\mathrm{axis}$ origin ($X_0=x_0/\sqrt{\pi}R_F$) is the intercept
of the source-step line with the observer's plane.
\label{diffpoint}}
\end{center}
\end{figure}

Fig. \ref{diffpoint} (left) displays the expected intensity variations in
the observer's plane
for a point-like monochromatic source observed through a transparent screen
with a step of optical path $\delta=\lambda/4$
and through a prism edge.
The inter-fringe is -- in a natural way --
close to the length scale defined by $\sqrt{\pi}R_F$.

Such configurations model the edge of a $\mathrm{H_2}$
structure, because they have the same average gradient of optical path.
More generally, contrasted patterns
take place as soon as the second derivative of the optical path
is different from zero.
%Discontinuity (of the optical path or of its
%derivatives, as in our examples) is not necessary to get such patterns.
\section{Limitations from spatial and temporal coherences}
At optical wavelengths, diffraction pattern contrast
is severely limited by the size of the source $r_S$.
Fig. \ref{diffpoint} (right) shows the diffraction
patterns for different reduced source radii, defined as
$R_S=r_s/(\sqrt{\pi}R_F)\times z_0/z_1$, where $z_1$ is the distance
from the source to the screen.
%The resulting contrast of the pattern is critically limited by the
%angular size of the source.
%Table \ref{configurations} gives the expected contrasts for several
%source positions and types.

%The standard UBVRI filter system has passbands that all satisfy
%$\Delta \lambda/\lambda <0.1$.
In return, temporal coherence with the standard UBVRI filters
is sufficiently high to enable the formation of contrasted interferences
in the configurations considered here.
The fringe jamming induced by the wavelength dispersion
is also much smaller than the jamming due to the source extension.
\section{What is to see?}
An interference pattern with inter-fringe
of $\sim R_F$ ($1000-10\,000\,\mathrm{km}$ at $\lambda=500\,\mathrm{nm}$) is expected
to sweep across the Earth when
the line of sight of a sufficiently small astrophysical source
crosses an inhomogeneous transparent Galactic
structure (see Table \ref{configurations}).
%that changes the optical path by a significant
%fraction of $\lambda$.
%Such structures move with respect to the line of sight with
%typical velocities given in table \ref{configurations}.
This pattern moves at the transverse velocity $V_T$ of the diffusive screen.
Its shape may also evolve, due to
random turbulence in the scattering medium.
%We will base the
%present study on the assumption that the scintillation is
%mainly due to pattern motion rather than the pattern instability,
%as it is usually the case in radioastronomy observations (\cite{lyne}).
For the Galactic $\mathrm{H_2}$ clouds we are interested in,
we expect a typical modulation index $m_{scint}$ (or contrast)
ranging between 1\% and $\sim 20\%$, critically depending on the source
apparent size.
The time scale $t_{scint} = R_F/V_T$ of the intensity variations
is of order of $30\,\mathrm{s}$.
As the inter-fringe scales with $\sqrt{\lambda}$,
one expects a significant difference in the time scale $t_{scint}$
between the red side of the optical
spectrum and the blue side. This property
might be used to sign the diffraction
phenomenon at the $R_F$ natural scale.
\begin{table*}
\begin{center}
{\small
\caption[]{Configurations leading to strong diffractive scintillation.
Here we assume a regime characterized by $R_{diff}\le R_F$,
or at least a transitory regime as described
in Section \ref{simplecase} -- characterized by $R_F$ --,
for example if
an inhomogeneity due to a turbulent mechanism crosses the line of sight.
Numbers are given for $\lambda=500\,\mathrm{nm}$.
}
\label{configurations}
\begin{tabular}{|c|c||c|c|c||c|c|c|}
\cline{3-8}
\multicolumn{2}{c||}{} & \multicolumn{6}{c|}{\bf SCREEN} \\
\cline{3-8}
\multicolumn{2}{c||}{} & atmos- & solar	& solar	& \multicolumn{3}{c|}{Galactic} \\
\multicolumn{2}{c||}{} & phere & system& suburbs  & thin disk& thick disk& halo \\
\hline
\multicolumn{2}{r||}{Distance}	& $10\,\mathrm{km}$	& $1\,\mathrm{AU}$ & $10\,\mathrm{pc}$	& $300\,\mathrm{pc}$ & $1\,\mathrm{kpc}$ & $10\,\mathrm{kpc}$ \\
\hline
\multicolumn{2}{r||}{$R_F$ to $\times$ by $\left[\frac{\lambda}{500\,\mathrm{nm}}\right]^{\frac{1}{2}}$} & $2.8\,\mathrm{cm}$	& $109\,\mathrm{m}$& $157\,\mathrm{km}$	& $860\,\mathrm{km}$& $1570\,\mathrm{km}$ & $5000\,\mathrm{km}$ \\
\hline
\multicolumn{2}{r||}{Transverse speed $V_T$}	& $1\,\mathrm{m/s}$	& $10\,\mathrm{km/s}$ & $20\,\mathrm{km/s}$	& $30\,\mathrm{km/s}$ & $40\,\mathrm{km/s}$ & $200\,\mathrm{km/s}$ \\
\hline
\multicolumn{2}{r||}{$t_{scint}$
to $\times\left[\frac{\lambda}{500\,\mathrm{nm}}\right]^{\frac{1}{2}}
\left[\frac{R_{diff}}{R_F}\right]$}	& $0.03\,\mathrm{s}$ & $0.01\,\mathrm{s}$ & $8\,\mathrm{s}$ & $29\,\mathrm{s}$ & $40\,\mathrm{s}$ & $25\,\mathrm{s}$ \\
\hline
%\multicolumn{2}{r||}{Optical depth $\tau_{scint}$} & $1$ & & & \multicolumn{3}{c|}{total $>10^{-7}$} \\
%\hline
\multicolumn{2}{r||}{$m_{scint}$ in \% to multiply by} & & $32\%\times$ & $2.2\%\times$ & $4.1\%\times$ & $2.2\%\times$ & $7.1\%\times$ \\
\multicolumn{2}{r||}{
$\left[\frac{\lambda}{500\,\mathrm{nm}}\right]^{\frac{1}{2}}
\left[\frac{R_{diff}}{R_F}\right]\left[\frac{r_S}{r_{\odot}}\right]^{-1}$
}	& $100\%$	& $\left[\frac{d}{10\mathrm{pc}}\right]$ & $\left[\frac{d}{1\mathrm{kpc}}\right]$	& $\left[\frac{d}{10\mathrm{kpc}}\right]$	& $\left[\frac{d}{10\mathrm{kpc}}\right]$ & $\left[\frac{d}{100\mathrm{kpc}}\right]$ \\
\hline
\hline
\multicolumn{2}{|c||}{\bf SOURCE} & \multicolumn{6}{c|}{\bf DIFFRACTIVE MODULATION INDEX $m_{scint}$} \\
\cline{1-2}
Location & Type & \multicolumn{6}{c|}{\bf (to multiply by $\sqrt{\lambda/500\,\mathrm{nm}}\times R_{diff}/R_F$)} \\
\hline
\cline{3-8}
nearby 10pc	& any star	& $\ll 1\%$ & $<100\%$ & & & & \\
\cline{1-2}\cline{4-8}
Galactic 8kpc & star & in a & 100\% & 1-70\% & 1-10\% &  &  \\
\cline{1-2}\cline{4-8}
LMC 55kpc & A5V ($r_S=1.7r_{\odot}$) & tele- & 100\% & 70\% & 13\% & 7\% & 2\% \\
\cline{1-2}\cline{4-8}
M31 725kpc & B0V ($r_S=7.4r_{\odot}$) & scope & 100\% & 100\% & 40\% & 22\% & 7\% \\
\cline{1-2}\cline{4-8}
z=0.2--0.9Gpc	& SNIa & $>1\,\mathrm{m}$ & 100\% & 70\% & 13\% & 7\% & 2\%\\
\cline{1-2}\cline{4-8}
z=1.7--1.7Gpc& Q2237+0305 & & 100\% & $>45\%$ & $>8\%$ & $>4\%$ & $>1.4\%$\\
\hline\end{tabular}
}
\end{center}
\end{table*}
\section{How to see it?}
{\bf Minimal hardware:}
from Table \ref{configurations}, we deduce that the minimal magnitude of
extragalactic stars that are likely to undergo a few percent modulation index
from a Galactic molecular cloud is $M_V\sim 20.5$ (A5V star in LMC
or B0V in M31).
Therefore, the search for diffractive scintillation
needs the capability to sample
every $\sim 10\,\mathrm{s}$ (or faster) the luminosity of stars with $M_V>20.5$,
with a point-to-point precision better than a few percent.
This performance can be achieved using a 2 meter telescope with a
high quantum efficiency detector allowing a negligible dead time
between exposures (like frame-transfer CCDs).
Multi-wavelength detection capability is highly desirable to
exploit the dependence of the diffractive scintillation pattern with
the wavelength.% $\lambda$.%as a signature of the process.

{\bf Chances to see something?}
The $1\%$ surface filling factor predicted
for gaseous structures % of the thick disk %that have a typical
is also the maximum optical depth for {\it all} the
possible refractive (weak or strong) and diffractive scintillation regimes.
%Nevertheless, we want to consider here the pessimistic case where the
%optical depth for the strong regimes is much smaller:
%The typical crossing time of a thick disk cloud is $\sim 400$ days.
Under the pessimistic hypothesis that strong diffractive regime occurs
only when a Galactic structure enters or leaves the line of sight,
the duration for this regime is of order of
$\sim 5$ minutes (time to cross a few fringes)
over a total typical crossing time of $\sim 400$ days.
Then the diffractive regime optical depth $\tau_{scint}$
should be at least of order of $10^{-7}$ and the average exposure
needed to observe one event of $\sim 5$ minute duration
is $10^{6}\,\mathrm{star\times hr}$ \footnote{
Turbulence or any process creating filaments,
cells, bubbles or fluffy structures should increase these estimates.}.
It follows that
a wide field detector is necessary to monitor a large number of stars.
\section{Foreground effects, background to the signal}
{\bf Atmospheric effects:}
Surprisingly, atmospheric intensity scintillation is negligible
through a large telescope ($m_{scint}\ll 1\%$ for
a $>1\,$m diameter telescope \cite{dravins}).
Any other long time scale atmospheric effect such as
absorption variations at the sub-minute scale
(due to fast cirruses for example) should be easy to recognize as long
as nearby stars are monitored together.

{\bf The solar neighbourhood:}
Overdensities at $10\,\mathrm{pc}$ could produce a signal very similar
to one expected from the Galactic clouds.
But in this case, even big stars should undergo
a contrasted diffractive scintillation;
the distinctive feature of scintillation through more distant
screens ($>300\,\mathrm{pc}$) is that {\it only the smallest stars}
are expected to scintillate.
%but only the smallest will scintillate through more distant
%screens ($>300\,\mathrm{pc}$).
It follows that simultaneous monitoring of
various types of stars at various distances
should allow one to discriminate effects due to solar neighbourhood
gas and due to more distant gaseous structures.

{\bf Sources of background?}
Physical processes such as asterosismology, granularity of the
stellar surface, spots or eruptions
produce variations of very different amplitudes and time scales.
A few categories of recurrent variable stars exhibit important emission
variations at the minute time scale \cite{sterken},
but their types are easy to identify from spectrum.
%UV-Ceti are also very faint stars (absolute magnitude $>15$) and
%only the closest ones could contaminate a monitoring sample.
\section{Conclusions and perspectives}
Structuration of matter is perceptible at all scales,
and the eventuality of stochastic fluctuations
producing diffractive scintillation is not rejected
by observations.
In this paper, I showed that there is an
observational opportunity resulting from
the subtle compromise between the arm-lever of
interference patterns due to hypothetic diffusive
objects in the Milky-Way and the size of the extra-galactic stars.
The hardware and software techniques required for scintillation searches
are currently available. Tests are under way to validate some of
the ideas discussed here.

If some indications are discovered with a single telescope, one will have
to consider a project involving a 2D array of telescopes, a few hundred
and/or thousand kilometers apart. Such a setup would allow to temporally
and spatially sample an interference pattern, unambiguously providing
the diffraction length scale $R_{diff}$, the speed and
the dynamics of the scattering medium.

\end{document}